\documentclass[journal]{IEEEtran}
\usepackage{amsmath,amsfonts}
\usepackage{algorithmic}
\usepackage{array}
\usepackage[caption=false,font=normalsize,labelfont=sf,textfont=sf]{subfig}
\usepackage{textcomp}
\usepackage{stfloats}
\usepackage{url}
\usepackage{graphicx}
\usepackage{balance}
\usepackage{orcidlink}
\usepackage{tikz}
\usetikzlibrary{arrows.meta, positioning, calc, fit, shapes.geometric, decorations.pathreplacing}
\begin{document}

\title{Interference-Graph-Aided Multi-Agent Reinforcement Learning for Beam Hopping LEO Satellite Networks}

\author{
    \IEEEauthorblockN{Hao Yuan\,\orcidlink{0000-0002-5670-6171}, Ziyi Yang\,\orcidlink{0009-0005-1569-1099}, Xing Zhang\,\orcidlink{0000-0003-4345-6166}, \textit{Senior Member, IEEE}}
\thanks{Manuscript received XXXX.}}


\markboth{IEEE Wireless Communications Letters}%
{Interference-Graph-Aided MARL for Beam Hopping LEO Satellite Networks}

\maketitle

\begin{abstract}
Beam hopping technology enables LEO satellites to dynamically switch beam coverage positions, providing flexible resource allocation for non-uniformly distributed traffic demands.
However, the severe spatial interference coupling among beams makes coordinated scheduling challenging.
To address this problem, in this letter, we propose an interference-graph-aided multi-agent reinforcement learning approach that constructs an interference graph to capture beam coupling relationships and injects this structural prior into agent-side representation learning.
Each beam agent aggregates information from interference neighbors via graph convolution before estimating local Q-values, which are then decomposed by the standard mixing network.
Simulation results demonstrate that the proposed method achieves 34.4\% higher throughput than QMIX with more stable interference management, under non-uniform traffic and co-channel interference settings.
\end{abstract}


\section{Introduction}
\label{sec:intro}
\IEEEPARstart{N}{ext-generation} satellite communication systems face the challenge of serving highly non-uniform and time-varying traffic demands across vast geographic areas.
Beam hopping technology addresses this by allowing satellite beams to dynamically switch coverage positions, concentrating resources where they are most needed~\cite{hu2019beam}.

Recent works have applied deep reinforcement learning (DRL) to beam hopping scheduling~\cite{du2022dynamic,wang2023interference}, demonstrating improved throughput and adaptability over conventional optimization methods.
However, these single-agent approaches treat the multi-beam system as a monolithic decision-maker, ignoring the distributed nature of beam coordination and scaling poorly with system size.

Multi-agent reinforcement learning (MARL) offers a natural framework for distributed beam coordination~\cite{marl_survey,dec_pomdp}.
QMIX~\cite{qmix} enables cooperative learning through value decomposition while maintaining decentralized execution.
Yet standard QMIX treats all agents uniformly, overlooking the critical insight that interference coupling varies significantly across beam pairs---strongly coupled beams require tighter coordination than weakly coupled ones.
Graph neural networks (GNNs) have emerged as powerful tools for capturing structural relationships in wireless networks~\cite{gnn_wireless2023}, and recent works combining GNNs with MARL~\cite{gnn_marl2025} demonstrate that graph-based agent representation can significantly improve coordination in structured multi-agent environments.

In this letter, we propose an interference-graph-aided MARL framework for beam hopping in multi-beam satellite systems.
Our key insight is that the interference topology provides a natural coordination prior: agents connected by strong interference should share more information during decision-making.
We construct an interference graph to capture beam coupling relationships and inject this structural prior into agent-side representation learning via graph convolution.
The main contributions are:
\begin{enumerate}
    \item A Dec-POMDP formulation for beam hopping in multi-beam satellite systems with interference-aware reward design.
    \item A graph-enhanced agent network that aggregates information from interference neighbors.
    \item Simulation results demonstrating 34.4\% throughput improvement over QMIX and more stable interference management.
\end{enumerate}

\section{System Model and Problem Formulation}
\label{sec:problem}

\subsection{System Model}
\label{sec:system}

Consider a satellite system with $N$ hopping beams serving $M$ ground cells.
At each time slot $t$, beam $i$ selects: (i) a hopping location $l_i^t \in \mathcal{L}_i$ from a discrete set of $L$ candidate positions, (ii) an activation decision $\beta_i^t \in \{0, 1\}$, and (iii) a power level index $\kappa_i^t \in \{1, \ldots, K\}$.
The actual transmit power is $P_i^t = \mathcal{P}[\kappa_i^t]$, where $\mathcal{P}[\cdot]$ maps the discrete index to a preset power value from $\{P_1, \ldots, P_K\}$ with $P_k \in [0, P_{\max}]$.

The achievable rate for cell $j$ at time $t$ is:
{\small\begin{equation}
\label{eq:rate}
R_j^t = \sum_{i=1}^{N} \beta_i^t \cdot \mathbb{1}\{l_i^t = j\} \cdot B \log_2\!\left(1 + \frac{P_i^t |h_{i,j}^t|^2}{\sum_{k \neq i} \beta_k^t P_k^t |h_{k,j}^t|^2 + \sigma^2}\right)
\end{equation}}
where $h_{i,j}^t$ is the channel gain from beam $i$ to cell $j$, following a distance-dependent model $h_{i,j} \propto d_{i,j}^{-1/2}$ with per-slot random perturbation, $B$ is the bandwidth, $\sigma^2$ is the noise power, and $\mathbb{1}\{\cdot\}$ is the indicator function.
Traffic demand $d_j^t$ is drawn from a uniform mixture $d_j = d_{\mathrm{base}} + d_{\mathrm{peak}} \cdot U(0,1)$ with spatial shuffling and per-slot fluctuation to model non-uniform traffic.

\begin{figure}[!t]
\centering
\includegraphics[width=\columnwidth]{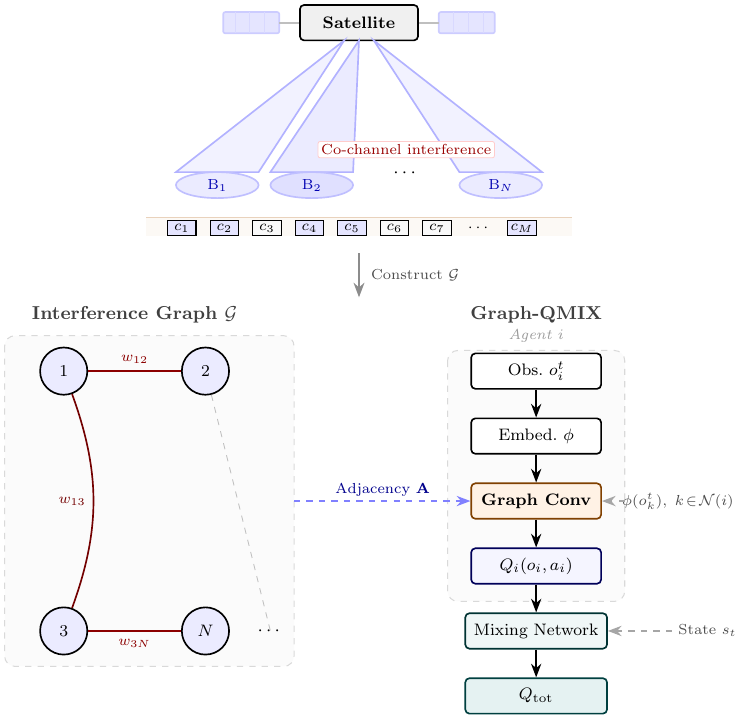}
\caption{System model of the proposed interference-graph-aided MARL for beam hopping satellites. The upper part illustrates the satellite beam hopping system with interference coupling. The lower-left shows the constructed interference graph $\mathcal{G}=(\mathcal{V},\mathcal{E},\mathbf{W})$. The lower-right depicts the Graph-QMIX architecture.}
\label{fig:system}
\end{figure}

\subsection{Dec-POMDP Formulation}
\label{sec:decpomdp}

We formulate the problem as a Dec-POMDP $(\mathcal{S}, \mathcal{A}, P, R, \Omega, O, \gamma)$ where each beam $i \in \{1,\ldots,N\}$ is an independent agent.
The global state $s_t \in \mathcal{S}$ includes traffic demands $\mathbf{d}^t \in \mathbb{R}^M$, channel gain matrix $\mathbf{H}^t \in \mathbb{R}^{N \times M}$, previous activation and location decisions, and the interference graph adjacency.
Each agent observes local demand in its coverage region, local channel quality, and neighbor activation states.
Each agent selects $a_i^t = (l_i^t, \beta_i^t, \kappa_i^t)$ as defined in Section~\ref{sec:system}, subject to $\sum_{i=1}^{N} \beta_i^t \leq N_{\max}$.

\textbf{Reward}: The global reward balances throughput, demand satisfaction, and interference penalty:
\begin{equation}
\label{eq:reward}
r_t = \omega_1 \underbrace{\sum_{j=1}^{M} R_j^t}_{\text{throughput}} + \omega_2 \underbrace{\frac{1}{M}\sum_{j=1}^{M} \mathbb{1}\{R_j^t \geq d_j^t\}}_{\text{demand satisfaction}} - \omega_3 \underbrace{C_{\mathrm{int}}^t}_{\text{interference}}
\end{equation}
where $\omega_1{=}1.0$, $\omega_2{=}1.0$, $\omega_3{=}3.0$ are weighting coefficients.
$C_{\mathrm{int}}^t$ counts co-channel beam pairs sharing the same location plus adjacent-location interference.

\textbf{Objective}: Maximize the expected cumulative discounted reward:
\begin{equation}
\label{eq:objective}
\max_{\pi} \mathbb{E}\left[\sum_{t=0}^{\infty} \gamma^t r_t \right]
\end{equation}
where $\gamma \in (0, 1)$ is the discount factor.

\section{Proposed Interference-Graph-Aided MARL}
\label{sec:method}

Standard QMIX learns cooperative policies but treats all agents uniformly, ignoring the interference topology.
In beam hopping systems, not all beam pairs have equal coupling strength---agents connected by strong interference should coordinate more closely.
We inject the interference structure into agent-side representation learning, while keeping value decomposition under the standard QMIX framework.
We construct an interference graph $\mathcal{G} = (\mathcal{V}, \mathcal{E}, \mathbf{W})$ to capture beam coupling relationships.
Each agent $i$ uses this graph to aggregate information from interference neighbors before estimating its Q-values.
The Q-values are then fed into the standard QMIX mixing network for global value decomposition. Since all beams originate from the same satellite platform, neighbor observation embeddings required by graph convolution can be exchanged through on-board shared memory with negligible overhead.

\subsection{Interference Graph Construction}
\label{sec:graph}

We construct an interference graph $\mathcal{G} = (\mathcal{V}, \mathcal{E}, \mathbf{W})$ where nodes represent beams and edge weights quantify coupling strength via three components:
\begin{equation}
\label{eq:weight}
w_{ik} = \alpha_1 \cdot \mathrm{corr}(\mathbf{h}_i, \mathbf{h}_k) + \alpha_2 \cdot p_{ik} + \alpha_3 \cdot s_{ik}
\end{equation}
where $\alpha_1{=}0.4$, $\alpha_2{=}0.3$, $\alpha_3{=}0.3$ are combining weights.
The \emph{channel correlation} term $\mathrm{corr}(\mathbf{h}_i, \mathbf{h}_k) \geq 0$ measures the similarity of channel gain profiles between beams $i$ and $k$.
The \emph{topology proximity} term $p_{ik} = \max(0, 1 - d_{ik}/(N/2))$ captures circular distance $d_{ik}$ on the beam ring.
The \emph{sidelobe coupling} term $s_{ik}$ is nonzero only for adjacent ($s{=}0.3$) and second-adjacent ($s{=}0.1$) beam pairs.
Edges with $w_{ik} > \tau$ are retained, where $\tau$ is the sparsification threshold.
The graph is reconstructed at each time step based on the evolving channel state.
We consider two variants for ablation: a \textbf{binary variant} where $A_{ik} \in \{0, 1\}$ indicates edge existence only, and the proposed \textbf{weighted variant} where $A_{ik} = w_{ik}$ preserves interference strength.

\subsection{Graph-Enhanced Agent Network}
\label{sec:agent}

Each agent employs a graph-enhanced network that incorporates interference-aware neighbor information into its local Q-value estimation.

\textbf{Graph Convolution}: We adopt a message-passing graph convolution layer inspired by~\cite{kipf2017}:
\begin{equation}
\label{eq:gnn}
\mathbf{z}_i = \varphi \left( \mathbf{W}_s \phi(o_i^t) + \sum_{k \in \mathcal{N}(i)} \tilde{A}_{ik} \mathbf{W}_n \phi(o_k^t) \right)
\end{equation}
where $\varphi(\cdot)$ is ReLU, $\mathcal{N}(i)$ denotes the neighbors of agent $i$, $\mathbf{W}_s$ and $\mathbf{W}_n$ are learnable weight matrices, $\phi(\cdot)$ is an MLP encoder mapping observation to embedding, and $\tilde{A}_{ik} = A_{ik} / \sum_j A_{ij}$ is the degree-normalized adjacency weight.

\textbf{Q-Network Output}: The graph-enhanced embedding $\mathbf{z}_i$ is fed into a per-agent Q-network:
\begin{equation}
\label{eq:qlocal}
Q_i(o_i, a_i) = f_Q(\mathbf{z}_i, a_i)
\end{equation}
where $f_Q$ is a three-layer MLP.

\subsection{Standard QMIX Mixing Network}
\label{sec:mixing}

We employ the standard QMIX mixing network~\cite{qmix}.
Given local Q-values $\{Q_1, \ldots, Q_N\}$, the mixing network produces $Q_{\text{tot}}$:
\begin{equation}
\label{eq:mixing}
Q_{\text{tot}} = f_{\text{mix}}(Q_1, \ldots, Q_N; s_t; \boldsymbol{\theta}_m)
\end{equation}
where $f_{\text{mix}}$ is an MLP with hypernetworks that take the global state $s_t$ as input.
The hypernetwork weights are passed through an absolute value activation to ensure monotonicity: $\frac{\partial Q_{\text{tot}}}{\partial Q_i} \geq 0, \forall i$, preserving IGM consistency.

\subsection{Training Procedure}
\label{sec:training}

The network is trained end-to-end by minimizing the temporal-difference (TD) loss:
\begin{equation}
\label{eq:loss}
\mathcal{L}(\theta) = \mathbb{E}_{(s, \mathbf{a}, r, s') \sim \mathcal{D}} \left[ \left( y - Q_{\text{tot}}(s, \mathbf{a}; \theta) \right)^2 \right]
\end{equation}
where $y = r + \gamma \max_{\mathbf{a}'} Q_{\text{tot}}(s', \mathbf{a}'; \theta^{-})$ is the target value, $\theta^{-}$ are the target network parameters, and $\mathcal{D}$ is the experience replay buffer.
We update the target network via soft Polyak averaging every 100 steps.
The exploration factor $\epsilon$ decays from 1.0 to 0.05 during training, encouraging early exploration and late exploitation.

\section{Simulation Results}
\label{sec:exp}

\subsection{Experimental Setup}

We evaluate the proposed method on a satellite system with $N{=}8$ hopping beams serving $M{=}32$ ground cells.
Each beam selects from $L{=}4$ hopping locations and $K{=}4$ discrete power levels, with at most $N_{\max}{=}3$ beams active simultaneously.
The interference graph sparsification threshold is $\tau = 0.05$.
All MARL agents are trained for 20,000 episodes (50 steps per episode) with learning rate $10^{-4}$, discount factor $\gamma{=}0.99$, replay buffer size $10^5$, and batch size 128.
The Q-network is a three-layer MLP with hidden dimension $d{=}128$, and the GCN uses shared weights with one graph convolution layer.
Target networks are updated via soft Polyak averaging ($\tau_{\mathrm{polyak}}{=}0.005$) every 100 steps, with $\epsilon$-greedy decay from 1.0 to 0.05.
Evaluation uses 100 test episodes with $\epsilon{=}0$, averaged over 3 random seeds (42, 123, 456).

\textbf{Baselines}: Fixed (static beam allocation), Random, Greedy (per-slot demand-based), IQL, VDN~\cite{vdn}, and QMIX~\cite{qmix}.

\subsection{Training Convergence}
\begin{figure}[!t]
\centering
\includegraphics[width=\columnwidth]{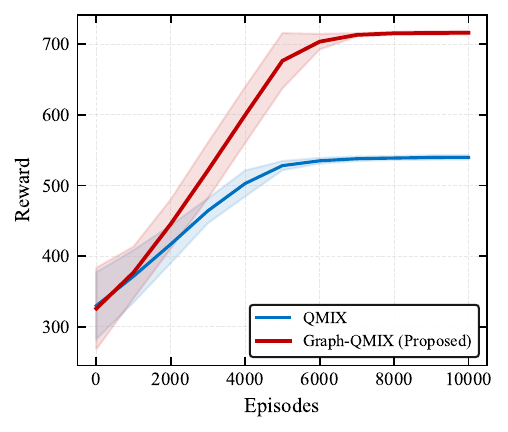}
\caption{Training convergence of QMIX and Graph-QMIX (3 seeds; shaded: $\pm$1 std).}
\label{fig:convergence}
\end{figure}

Fig.~\ref{fig:convergence} shows the training convergence of QMIX and Graph-QMIX.
Both converge within approximately 5,000 episodes, but Graph-QMIX achieves a significantly higher converged reward (${\sim}716$ vs.\ ${\sim}541$, a 32.4\% improvement), with shaded bands confirming stability across seeds.

\subsection{Throughput Comparison}
\label{sec:comparison}

\begin{figure}[!t]
\centering
\includegraphics[width=\columnwidth]{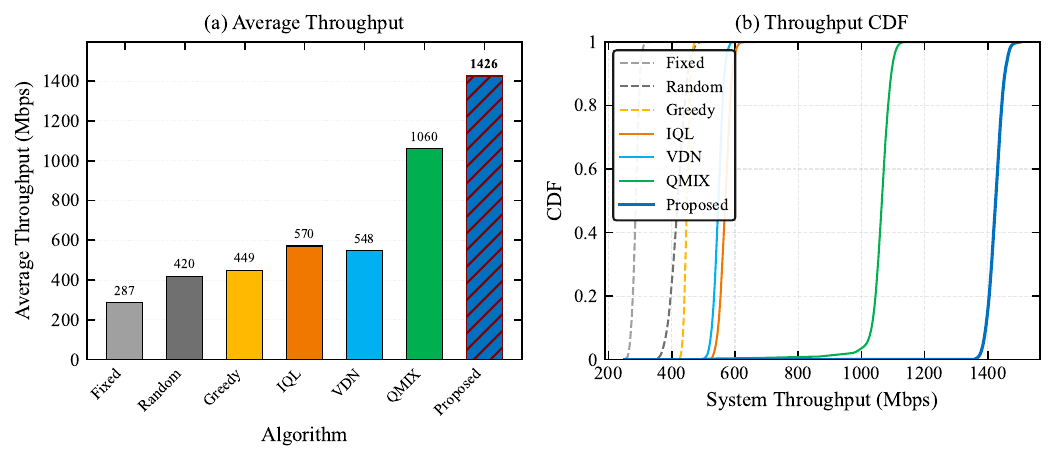}
\caption{Throughput comparison across algorithms.}
\label{fig:throughput}
\end{figure}

Fig.~\ref{fig:throughput} and Table~\ref{tab:main} present the throughput comparison.
The proposed Graph-QMIX achieves 1,425.6 Mbps average throughput, outperforming QMIX by 34.4\%.
The improvement stems from the graph-enhanced agent network enabling interference-aware coordination among beams.
Notably, MARL-based methods significantly outperform heuristic approaches, demonstrating the advantage of learning-based scheduling.
Among MARL methods, the performance gap widens as coordination complexity increases: IQL and VDN achieve only 570.5 and 548.3 Mbps respectively, while QMIX with monotonic value decomposition reaches 1,060.3 Mbps.
The proposed graph-enhanced extension further boosts throughput by incorporating interference topology into agent representation.
The proposed method also achieves 98.4\% demand satisfaction rate (DSR), significantly higher than QMIX's 73.2\%.

\begin{table}[t]
\centering
\caption{Performance Comparison}
\label{tab:main}
\begin{tabular}{|l|c|c|c|}
\hline
\textbf{Method} & \textbf{Tput (Mbps)} & \textbf{SINR (dB)} & \textbf{DSR (\%)} \\
\hline
Fixed       & $287.3 \pm 11.0$  & $4.83 \pm 0.38$ & $21.6$ \\
Random      & $419.6 \pm 25.1$  & $5.14 \pm 0.31$ & $32.7$ \\
Greedy      & $449.0 \pm 8.9$   & $8.09 \pm 0.17$ & $33.9$ \\
IQL         & $570.5 \pm 17.3$  & $7.36 \pm 0.32$ & $45.1$ \\
VDN         & $548.3 \pm 15.6$  & $10.98 \pm 0.38$ & $41.4$ \\
QMIX        & $1060.3 \pm 51.1$ & $13.52 \pm 0.36$ & $73.2$ \\
\textbf{Proposed} & $\mathbf{1425.6 \pm 23.0}$ & $\mathbf{13.56 \pm 0.22}$ & $\mathbf{98.4}$ \\
\hline
\end{tabular}
\end{table}

\subsection{Interference Analysis}

\begin{figure}[!t]
\centering
\includegraphics[width=\columnwidth]{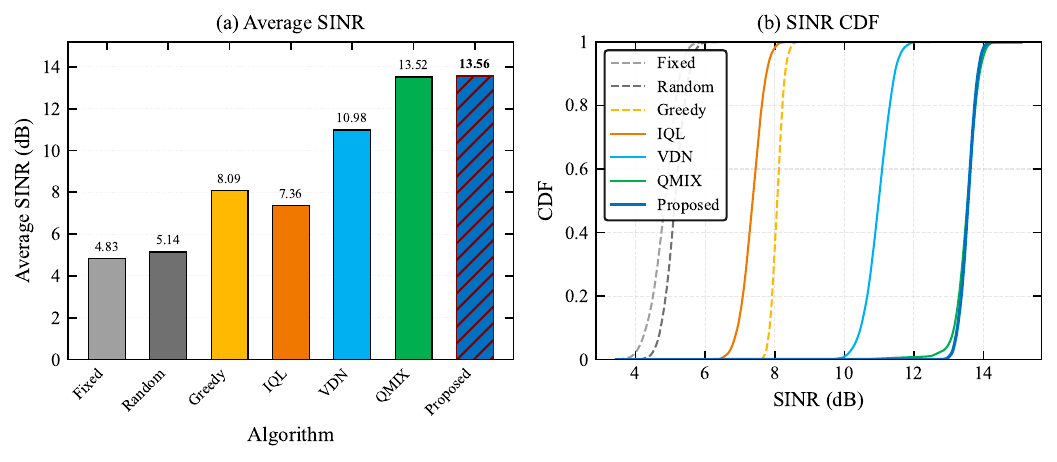}
\caption{SINR comparison. Left: average SINR with standard deviation. Right: CDF of per-episode SINR.}
\label{fig:interference}
\end{figure}

Fig.~\ref{fig:interference} compares the SINR across algorithms.
While the proposed method achieves a comparable mean SINR to QMIX (13.56 vs.\ 13.52 dB), its standard deviation is 39\% lower (0.22 vs.\ 0.36 dB), indicating more stable interference management.
This stability arises because the graph-enhanced agents explicitly model interference coupling, leading to more consistent coordination decisions.
Removing the graph component reduces throughput by 34.4\%, confirming that the interference-aware representation is the primary driver of the performance gain.

\subsection{Ablation Study}

Table~\ref{tab:ablation} presents the ablation results under the default non-uniform traffic.
Adding the interference graph (QMIX $\to$ Binary Graph) yields a 33.9\% throughput gain, confirming the value of graph-based coordination.
Using weighted edges (Binary $\to$ Proposed) provides a marginal further improvement, indicating that the primary benefit stems from the graph topology---identifying which beams should coordinate---rather than the precise weight magnitudes.

\begin{table}[t]
\centering
\caption{Ablation Study Results}
\label{tab:ablation}
\begin{tabular}{|l|c|}
\hline
\textbf{Variant} & \textbf{Throughput (Mbps)} \\
\hline
Proposed (weighted graph) & $\mathbf{1425.6 \pm 23.0}$ \\
Binary Graph & $1420.0 \pm 24.8$ \\
w/o Graph (QMIX) & $1060.3 \pm 51.1$ \\
\hline
\end{tabular}
\end{table}

\subsection{Parameter Sensitivity}

\begin{figure}[!t]
\centering
\includegraphics[width=\columnwidth]{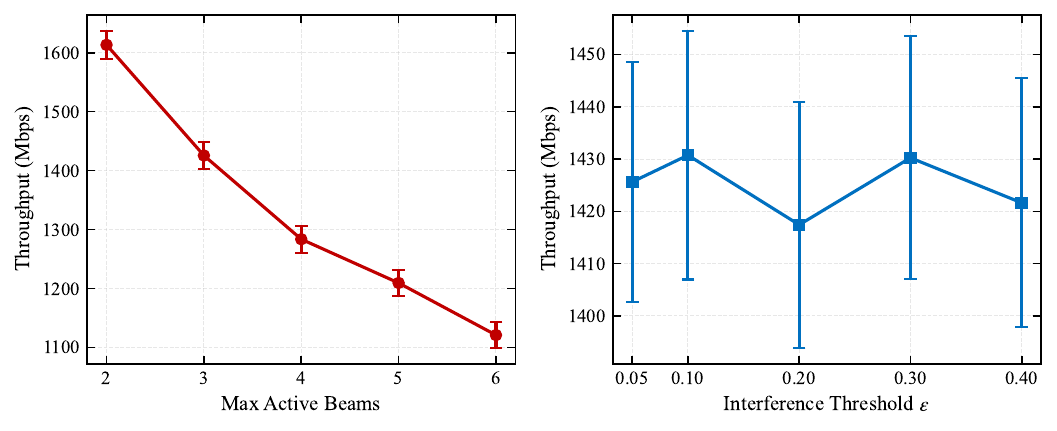}
\caption{Parameter sensitivity. (a) Impact of maximum active beams $N_{\max}$. (b) Impact of interference threshold $\tau$.}
\label{fig:sensitivity}
\end{figure}

Fig.~\ref{fig:sensitivity} evaluates the proposed method under varying system parameters.

\textbf{Impact of $N_{\max}$}: As the maximum number of active beams increases from 2 to 6, throughput decreases from 1613.3 to 1121.0 Mbps.
This is expected: more simultaneously active beams cause stronger co-channel interference, reducing per-beam rates.
The proposed method maintains good performance across all configurations, demonstrating its adaptability to varying interference conditions.

\textbf{Impact of $\tau$}: Varying the interference threshold from 0.05 to 0.4 changes throughput by less than 1.5\% (range: 1417.4--1430.7 Mbps), demonstrating that the method is robust to graph construction parameters.
A sparse graph ($\tau = 0.4$, fewer edges) performs comparably to a dense graph ($\tau = 0.05$, more edges), indicating that the graph convolution effectively filters noise from weakly coupled beam pairs without sacrificing coordination quality.

\section{Conclusion}
\label{sec:conclusion}

We proposed an interference-graph-aided MARL approach for beam hopping satellites.
By constructing an interference graph and injecting this structural prior into agent-side representation learning, our method enables more effective coordination among strongly coupled beams.
Simulation results show that the proposed method achieves 1,425.6 Mbps average throughput, a 34.4\% improvement over standard QMIX, while maintaining more stable interference management (SINR standard deviation reduced by 39\%).
Ablation studies confirm that the graph topology is the primary driver of performance gains, while the method achieves 98.4\% demand satisfaction rate.
The method demonstrates robust performance across different numbers of active beams and interference thresholds.
Future work includes extending to multi-satellite scenarios with dynamic topology changes and incorporating traffic prediction for proactive resource allocation.

\bibliographystyle{IEEEtran}

\begin{thebibliography}{99}

\bibitem{hu2019beam} X. Hu, S. Liu, Y. Wang, L. Xu, Y. Zhang, C. Wang, et al., ``Deep reinforcement learning-based beam hopping algorithm in multibeam satellite systems,'' IET Commun., vol. 13, no. 16, pp.~2485--2491, 2019. 10.1049/iet-com.2018.5774

\bibitem{du2022dynamic} X. Du, X. Hu, Y. Wang, and W. Wang, ``Dynamic resource allocation for beam hopping satellites communication system: An exploration,'' in Proc. IEEE Int. Conf. Trust, Security and Privacy in Computing and Communications (TrustCom), 2022, pp. 1296--1301. 10.1109/TrustCom56396.2022.00181

\bibitem{wang2023interference} H. Wang, L. Liu, X. Zhou, L. Xu, G. Wu, and S. Liu, ``Deep reinforcement learning based interference avoidance beam-hopping allocation algorithm in multi-beam satellite systems,'' in Proc. IEEE Int. Conf. Trust, Security and Privacy in Computing and Communications (TrustCom), 2023, pp. 1966--1973. 10.1109/TrustCom60117.2023.00268

\bibitem{marl_survey} K. Zhang, Z. Yang, and T. Ba\c{s}ar, ``Multi-agent reinforcement learning: A selective overview of theories and algorithms,'' in Handbook of Reinforcement Learning and Control. Cham, Switzerland: Springer, 2021, pp. 321--384.

\bibitem{dec_pomdp} D. S. Bernstein, R. Givan, N. Immerman, and S. Zilberstein, ``The complexity of decentralized control of Markov decision processes,'' Math. Oper. Res., vol. 27, no. 4, pp.~819--840, 2002. 10.1287/moor.27.4.819.297

\bibitem{qmix} T. Rashid, M. Samvelyan, C. S. de~Witt, G. Farquhar, J. Foerster, and S. Whiteson, ``Monotonic value function factorisation for deep multi-agent reinforcement learning,'' J. Mach. Learn. Res., vol. 21, no. 178, pp.~1--51, 2020.

\bibitem{gnn_wireless2023} Y. Shen, J. Zhang, S. H. Song, and K. B. Letaief, ``Graph neural networks for wireless communications: From theory to practice,'' IEEE Trans. Wirel. Commun., vol. 22, no. 5, pp.~3554--3569, May 2023. 10.1109/TWC.2022.3219840

\bibitem{gnn_marl2025} Y. Choi, P. G. Di~Marco, and P. Park, ``Communication-aware graph neural network for multi-agent reinforcement learning,'' IEEE Access, vol. 13, pp.~55832--55840, 2025. 10.1109/ACCESS.2025.3554736

\bibitem{kipf2017} T. N. Kipf and M. Welling, ``Semi-supervised classification with graph convolutional networks,'' in Proc. Int. Conf. Learn. Represent. (ICLR), 2017.

\bibitem{vdn} P. Sunehag, G. Lever, A. Gruslys, W. M. Czarnecki, V. Zambaldi, M. Jaderberg, et al., ``Value-decomposition networks for cooperative multi-agent learning based on team reward,'' in Proc. Int. Conf. Auton. Agents Multiagent Syst. (AAMAS), 2018, pp. 2085--2087. 10.65109/JSRC7365

\end{thebibliography}

\end{document}